\documentclass[]{bytedance_seed}

\usepackage[toc,page,header]{appendix}

\usepackage{tcolorbox}
\tcbuselibrary{skins} 

\usepackage{minitoc}
\usepackage{tablefootnote}
\usepackage[dvipsnames]{xcolor}
\usepackage[most]{tcolorbox}
\usepackage{upquote}
\title{SWE-Mirror: Scaling Issue-Resolving Datasets by Mirroring Issues Across
Repositories}

\author[1,2,*]{Junhao Wang}
\author[1, \dagger]{Daoguang Zan}
\author[1]{Shulin Xin}
\author[1]{Siyao Liu}
\author[1]{Yurong Wu}
\author[1, \dagger]{\mbox{Kai Shen}}

\affiliation[1]{ByteDance Seed}
\affiliation[2]{The Chinese University of Hong Kong}

\contribution[*]{Work done at ByteDance Seed}
\contribution[\dagger]{Corresponding authors}

\usepackage{xspace}
\usepackage{xparse}
\usepackage{pifont}

\newcommand{\swemirror} {\textsc{SWE-Mirror}}
\newcommand{\dataset} {\textsc{SWE-Mirro-60K}}

\newcommand{\testpatch} {test.patch}
\newcommand{\fixpatch} {fix.patch}
\newcommand{\taskpatch} {mirror.patch}

\newcommand{\phaseone} {Task Collection}
\newcommand{\phasetwo} {Task Mirroring}
\newcommand{\phasethree} {Task Verification}

\ExplSyntaxOn
\NewDocumentCommand{\trans}{m}{
  \texttt{%
    \clist_set:Nn \l_tmpa_clist { #1 }
    \clist_use:Nn \l_tmpa_clist { \ensuremath{\rightarrow} }%
  }%
}
\ExplSyntaxOff

\definecolor{mygreen}{rgb}{0,0.6,0}
\definecolor{myred}{rgb}{0.8,0,0}
\definecolor{myorange}{rgb}{1.0, 0.65, 0.0}

\newcommand{\cmark}{\textcolor{mygreen}{\ding{51}}}
\newcommand{\xmark}{\textcolor{myred}{\ding{55}}}
\newcommand{\partialcmark}{\textcolor{myorange}{\ding{51}}}

\definecolor{rq-bg}{RGB}{230, 250, 230}
\definecolor{rq-border}{RGB}{90, 87, 83}

\abstract{
Creating large-scale verifiable training datasets for issue-resolving tasks is a critical yet notoriously difficult challenge. Existing methods on automating the Gym environment setup process for real-world issues suffer from low success rates and high overhead. Meanwhile, synthesizing new tasks within existing Gym environments leaves the vast pool of authentic, human-reported problems untapped. To maximize the utilization of existing Gym environments and also the rich data of issue-resolving history on GitHub, we introduce \swemirror{}, a pipeline that distills a real-world issue’s semantic essence, mirrors it into another repository with a configured Gym environment, and re-animates it as a verifiable issue-resolving task. \swemirror{} reuses existing Gym environments along with the vast pool of issue-resolving history hosted on GitHub to construct a large-scale dataset of \textit{mirrored} authentic and verifiable tasks. Applying \swemirror{} to 40 repositories across 4 languages, we have curated a dataset with 60,671 issue-resolving tasks and demonstrated the value of our dataset by training and evaluating coding agents at various scale. Post-training experiments show that models trained with the dataset exhibit improvements in issue-resolving capabilities. Furthermore, by extending the dataset size to over 12,000 high-quality trajectories, we established a new state-of-the-art (SOTA) among Qwen2.5-Coder-Instruct based LLMs on the OpenHands agent framework, which increases the resolve rate on SWE-Bench-Verified by \textbf{+21.8\%} for the 7B model and \textbf{+46.0\%} for the 32B model and validates the effectiveness of our approach.

}

\date{\today}
\correspondence{\email{zandaoguang@bytedance.com} and \email{shen.kai@bytedance.com}}

\begin{document}

\maketitle

\abstract{
Creating large-scale verifiable training datasets for issue-resolving tasks is a critical yet notoriously difficult challenge. Existing methods on automating the Gym environment setup process for real-world issues suffer from low success rates and high overhead. Meanwhile, synthesizing new tasks within existing Gym environments leaves the vast pool of authentic, human-reported problems untapped. To maximize the utilization of existing Gym environments and also the rich data of issue-resolving history on GitHub, we introduce \swemirror{}, a pipeline that distills a real-world issue’s semantic essence, mirrors it into another repository with a configured Gym environment, and re-animates it as a verifiable issue-resolving task. \swemirror{} reuses existing Gym environments along with the vast pool of issue-resolving history hosted on GitHub to construct a large-scale dataset of \textit{mirrored} authentic and verifiable tasks. Applying \swemirror{} to 40 repositories across 4 languages, we have curated a dataset with 60,671 issue-resolving tasks and demonstrated the value of our dataset by training and evaluating coding agents at various scale. Post-training experiments show that models trained with the dataset exhibit improvements in issue-resolving capabilities. Furthermore, by extending the dataset size to over 12,000 high-quality trajectories, we established a new state-of-the-art (SOTA) among Qwen2.5-Coder-Instruct based LLMs on the OpenHands agent framework, which increases the resolve rate on SWE-Bench-Verified by \textbf{+21.8\%} for the 7B model and \textbf{+46.0\%} for the 32B model and validates the effectiveness of our approach.

}

\section{Introduction}
\newtcolorbox{researchquestionbox}[1][]{
    enhanced,
    colback=rq-bg,
    colframe=rq-border,
    boxrule=0.6pt, 
    arc=2mm,           
    fontupper=\itshape, 
    halign=left,
    boxsep=5pt,
    top=5pt,
    bottom=5pt,
    #1 
}


Large Language Models (LLMs) have demonstrated remarkable capabilities in various code generation tasks~\citep{chen2021humaneval, austin2021mbpp, evalplus, evalperf, jain2024livecodebench, alphacode, luo2025wizardcoderempoweringcodelarge, guo2024deepseekcoderlargelanguagemodel, wan2024divideconquer}, fundamentally reshaping the landscape of software development. As the research community broadens its focus to more complex and real-world challenges~\citep{zhang2024cutcrap, RUC2024agent-memory, jiang2025screencoder}, resolving real-world issues has emerged as a critical frontier~\citep{jimenez2024swebench, openai2024swebenchverified, zan2025msb, wei2025swerl}. A verifiable issue-resolving task instance, exemplified by benchmarks like SWE-Bench~\citep{jimenez2024swebench, openai2024swebenchverified, yang2025swesmith}, consists of two primary components:
\begin{itemize}
    \item \textbf{Task Context}: This includes the issue with related pull-request(\textit{i.e.}, PR) and the corresponding repository snapshot(\textit{i.e.}, CodeBase). Normally we can get a problem statement detailing a specific issue (\textit{e.g.}, a bug report or feature request) as the task description, and reference patches for validation and groundtruth. 
    \item \textbf{Gym}: This is an executable environment equipped with validation harness, including test commands and log parsers which can verify the correctness of proposed solutions and provide reward for training.
\end{itemize}

A severe imbalance exists~\cite{pan2025swegym, swerebench} in the effort required to acquire these two components. While Task Contexts can be gathered from platforms like GitHub with relative ease, engineering a functional Gym is a significant bottleneck, demanding meticulous and often unscalable manual effort~\citep{jimenez2024swebench, zan2025msb, pan2025swegym}. This difficulty arises because a universal, one-size-fits-all Gym is infeasible in the diverse software ecosystem. Each repository—and often, each specific version—requires a unique configuration of dependencies, build processes, and testing frameworks. Consequently, the immense effort invested in creating a single Gym typically supports only one specific task or, at best, a small cluster of closely related ones. This reality forges a rigid \textbf{one-to-one dependency} between Task Context and Gym, posing a fundamental barrier to scaling up the issue-resolving datasets.

Faced with this scaling challenge, the research community has pursued two orthogonal approaches to scaling the issue-resolving dataset for training: \ding{182} \textbf{Scaling tasks via synthesizing problems.} This approach maximizes the utility of Gyms by synthesizing new tasks that are compatible with them. Works like SWE-smith~\citep{yang2025swesmith} and SWE-Synth~\citep{pham2025swesynth} programmatically mutate or rewrite repositorys' components to inject bugs and generate a large volume of artificial tasks. \ding{183} \textbf{Scaling tasks via setting up Gyms.} This orthogonal approach confronts the Gym creation bottleneck directly by attempting to automate the environment setup process~\citep{swerebench}.

While both approaches offer paths to scale, they present a difficult trade-off. The synthesis approach achieves scale but generates problems that are artificially created, failing to leverage the vast and rich history of authentic software evolution found on platforms like GitHub—the very source of problems this research field aims to solve. Conversely, the Gym automation approach engages with this real-world data but faces significant engineering hurdles. The success rate of automatically configuring stable environments remains low, and the method incurs staggering storage costs. With each Gym environment consuming approximately 1GB, scaling to 100,000 instances would demand a 100 Terabytes of storage.

This presents the community with an untenable choice: pursue scalability with tasks disconnected from rich source of real-world software evolution, or engage with authentic data at a prohibitive engineering and storage cost. This dilemma leads to a research question:

\begin{researchquestionbox}
    How can we leverage the vast and ever-growing history of software evolution on GitHub using only a small, manageable set of reusable Gyms?
\end{researchquestionbox}

To answer this question, we must break the \textbf{one-to-one dependency} between the Task Context and the Gym. Our approach involves hosting an issue-resolving task from one repository within a pre-existing Gym configured for another. We draw inspiration from research on \textbf{issue mirroring}~\citep{guan2025crossprobe}, which observes that programs with analogous functionalities often share analogous bugs and features. While prior work has leveraged this insight to \textit{find} bugs across similar frameworks (\textit{e.g.}, PyTorch\footnote{\url{https://pytorch.org}} and TensorFlow\footnote{\url{https://www.tensorflow.org}}), we propose to significantly extend this idea to programmatically \textit{mirror} them—re-instantiating a PR from a source project into a target project to create a new task. Observations supporting the feasibilty can be summarized as follows:

\begin{enumerate}
    \item \textit{Shared Analogous Components:} Functionally similar projects often share analogous components rooted in common architectural patterns, dependencies, or high-level APIs and may suffer similar issues.
    \item \textit{Portable Problem Logic:} Software issues often encapsulates a core logical problem that can be abstracted from its original implementation and be re-instantiated as a new task within a similar project.
    \item \textit{Transferable Validation:} Issues from a source repository is typically accompanied by a validation mechanism (\textit{e.g.}, a test case that fails before the fix and passes after). which can be adapted and transferred to the target repository to verify the successful replication of the issue.
\end{enumerate}
To this end, we introduce \textsc{\swemirror{}}, a pipeline that systematically mirrors real-world PRs and issues from a source repository in the wild into a functionally similar target repository which has a configured Gym. By breaking the \textbf{one-to-one dependency} between Task Context and Gym, \textsc{\swemirror{}} dramatically multiplies the availible tasks of any single Gym and unlocking a vast pool of authentic issue-resolving histories. The main contributions of this paper are summarized as follows:
\begin{itemize}
    \item \textbf{Technique:} We propose \textsc{\swemirror{}}, a novel paradigm and a concrete methodology for scaling issue-resolving datasets by mirroring real-world issues into configured Gyms across repository boundaries.
    \item \textbf{Large-Scale Dataset:} We release \dataset{}, a large-scale dataset containing over 60,000 verifiable tasks. These tasks are composed of authentic issues mirrored into a small set of robust Gyms. A comparison with other datasets is shown in \Cref{tab:dataset_comparison}.
    \item \textbf{Empirical Validation and Methodology:} We conduct extensive experiments exploring various agentic posttraining methods on \dataset{}. Our results not only demonstrate that finetuned models achieve significant performance gains on standard benchmarks like SWE-bench Verified~\citep{openai2024swebenchverified} and \mbox{Multi-SWE-bench-Flash~\citep{zan2025msb}}, but also provide insights into training strategies for this domain. Also, we provide strong empirical evidence for scaling law~\citep{kaplan2020scalinglaw} of dataset size in the software engineering context.
\end{itemize}

\begin{table*}[t!]
\centering
\caption{Comparison of \swemirror{} with other issue-resolving datasets. The symbols indicate whether a dataset possesses the feature (\cmark{}), lacks it (\xmark{}), or possesses it partially (\partialcmark{}).}
\label{tab:dataset_comparison} 
\begin{tabular}{l| c| c| c| c| c}
\toprule
\textbf{Dataset} & \textbf{\#Tasks} & \textbf{\# Repos} & \textbf{Hidden Tests?} & \textbf{Verifiable?} &  \textbf{Env. Storage} \\
\midrule
SWE-rebench~\citep{swerebench} & 20k & 2k & \cmark{} & \cmark{} & - \\
SWE-Gym~\citep{pan2025swegym} & 2.4k & 11 & \cmark{} & \cmark{} & 6 TBs \\
SWE-Fixer~\citep{xie2025swefixer} & 110k &  856 & \partialcmark{} & \xmark{} & - \\
SWE-Smith~\citep{yang2025swesmith} & 50k & 128 & \xmark{} & \cmark{} & 295 GBs \\
\midrule
\textbf{\dataset{} (Ours)} & \textbf{60k} & 40  & \cmark{} & \cmark{} & 100GBs \\
\bottomrule
\end{tabular}
\end{table*}

\section{Scaling Issue-Resolving Tasks with \textsc{\swemirror{}}}

\begin{figure*}[htbp]
    \centering
    \includegraphics[width=\linewidth]{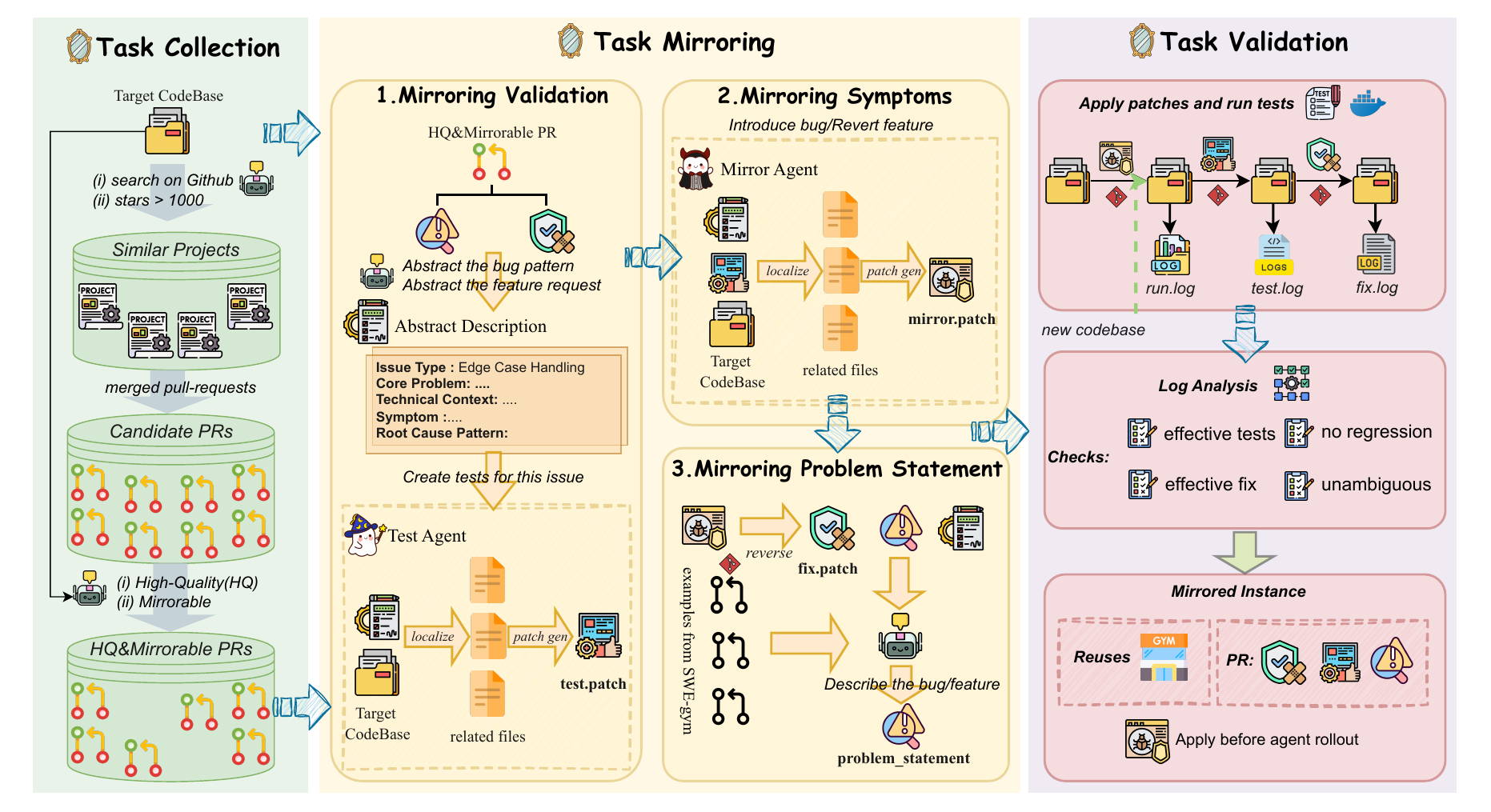}
    \caption{Overview of \swemirror{} pipeline.}
    \label{fig:main}
\end{figure*}

As illustrated in \Cref{fig:main}, this process is structured as a three-phase pipeline: \textit{(1) \phaseone}, where we collect high-quality and mirrorable real-world issues from GitHub; \textit{(2) \phasetwo}, where we mirror these issues into target codebases; and \textit{(3) \phasethree}, which validates the integrity of the mirrored task instances. Worthnoting, \swemirror{} is an orthoganal method on scaling dataset to prior efforts working on setting up Gyms for SWE instances. For seed Gym selection, due to the limit of resources and time, we select Gyms for newest issue from SWE-Gym~\cite{pan2025swegym}, SWE-rebench~\cite{swerebench} and Multi-SWE-RL~\cite{zan2025msb}, and set the time limit of running the \textit{whole} test suites to 5 minutes and the memory limit to 1GB. In addition, we also perform basic functional check of each Gym via running all testsuites and check the output manually.

\subsection{Phase 1: \phaseone}\label{ssec:task_collection}
The objective of this initial phase is to source a pool of potentially mirrorable issues for each target CodeBase with existing Gym. Given the vast volume of issues on GitHub, we employ a two-stage search strategy to narrow the candidate pool to a manageable scope. For a given CodeBase, we first leverage \texttt{Qwen3-32B}~\citep{yang2025qwen3technicalreport} to analyze its README file and generate five descriptive keywords. Using the GitHub REST API\footnote{\url{https://docs.github.com/en/rest}}, we then search for repositories using these keywords as query, retrieving the top 20 repositories ranked by stars and issue counts. Subsequently, we collect all pull-requests and linked issues from these candidate repositories and apply a filtering process, using a combination of hand-crafted rules and LM-based heuristic to identify \textit{high-quality} and \textit{mirrorable} issues. We expand the rules and LM-based heuristic in \Cref{appdix:pr_collection}.

\subsection{Phase 2: \phasetwo}\label{ssec:issue_mirroring}
The objective of this phase is to mirror the candidate issues into their designated target Gyms. The process begins by employing \texttt{GPT-4o-2024-0513}~\citep{gpt4o} to distill the related functionality, core logic, current and expected behavior and observable symptoms of a source issue into a concise \textit{abstract description} which serves as a primary input for our three-step mirroring workflow with \texttt{GPT-4.1}~\citep{gpt4.1} as the backbone LM:
\begin{itemize}
\item \textbf{Mirroring Validation}: The primary goal of this initial step is to establish a concrete, executable contract that formally defines what constitutes a correct resolution of the issue-resolving task. An agent refered as Test Agent, prompted with the \textit{abstract description}, is responsible for generate a new test case within the target Gym's existing test suite. Those tests are designed to \textit{pass} under the current codebase state, but will \textit{fail} once the next step introduced the issue successfully. The output of this step is the \texttt{\testpatch{}}. This patch serves a dual purpose: it acts as a precise guide for the next step and, ultimately, as the hidden tests for evaluating the correctness of submissions from coding agents.

\item \textbf{Mirroring Symptom}: With the validation tests established by the \texttt{\testpatch{}}, this second step aims to \textit{re-animate} the issue within the target repository's code. A different agent(Mirror Agent) takes the \textit{abstract description} for semantic context and the file paths and function names from the \texttt{\testpatch{}} as a strong structural prior. Its objective is to surgically modify the application's source code to specifically cause the new test case to fail. The resulting modification, packaged as the \texttt{\taskpatch{}} after removing comments, becomes the starting point of the issue-resolving task. We also programmatically create its inverse, the \texttt{\fixpatch{}}, which serves as the ground-truth solution.

\item \textbf{Mirroring Problem Statement}: This final step is responsible for synthesizing a natural-language problem description that will be presented to the coding agents. The goal is to create a description that is not only accurate to the mirrored bug but also feels native. To achieve this, the LM is prompted with a rich set of context following the quality criteria described in SWE-bench-Verified~\cite{openai2024swebenchverified}, including: (1) the original GitHub issue description for semantic context; (2) the generated \texttt{\testpatch{}} and the \texttt{\fixpatch{}} to ground the description in the specific files and functions of the target codebase; and (3) few-shot examples of other issues from the SWE-Gym to ensure stylistic consistency. The resulting \texttt{problem\_statement} synthesizes these inputs into a coherent, self-contained task description.
\end{itemize}

The successful execution of this workflow yields a final mirrored task. Each task instance is a self-contained data structure containing the following fields:
\begin{itemize}
\item \texttt{\taskpatch{}}: A patch that introduces a bug or reverts a feature in the codebase. Applying this patch creates the starting point of the issue-resolving task.
\item \texttt{\testpatch{}}: A patch used to test the correctness of a submission, in line with benchmarks like (Multi-)SWE-Bench~\cite{jimenez2024swebench, zan2025msb, openai2024swebenchverified}. This should not be revealed to the coding agent system.
\item \texttt{\fixpatch{}}: The groundtruth solution for the task, created by simply reversing the \texttt{\taskpatch{}}.
\item \texttt{problem\_statement}: The natural-language task description that is presented to the coding agents.
\end{itemize}
Detailed workflow design and prompts used in this phase are demonstrated in \Cref{appdix:workflow}.

\subsection{Phase 3: Task Verification}\label{ssec:issue_verify}
In this Phase, we first perform a sanity check to ensure all patches can be applied without error. Concretely,  the \texttt{\taskpatch{}} can be applied to the \texttt{base\_commit} of the original code base, the \texttt{\testpatch{}} and \texttt{\fixpatch{}} should be appliable after the application of \texttt{\taskpatch{}}. Then we conduct
an execution-based validation, executing the full test suite under three states.
\begin{enumerate}
    \item \texttt{Run.log}: Run all tests after apply \texttt{\taskpatch{}}.
    \item \texttt{Test.log}: Run all tests after apply \texttt{\taskpatch{}} and \texttt{\testpatch{}}.
    \item \texttt{Fix.log}: Run all tests after apply all three patches.
\end{enumerate} 

Following Multi-SWE-bench~\cite{zan2025msb}, we analyze the test status transitions across these logs and apply strict filtering rules to accept only unambiguously correct mirrored tasks:
\begin{enumerate}
    \item \textbf{Effective Tests: } the application \texttt{test.patch} should introducing new tests whithout affecting original tests. Comparing test status in \texttt{Run.log} and \texttt{Test.log}. Only \trans{PASSED, PASSED}, \trans{FAILED, FAILED} \trans{SKIPED, SKIPED}, and \trans{NONE, FAILED} are permitted.
    
    \item \textbf{Effective Fix:} The \texttt{fix.patch} must fixes somethings. So comparing status in three logs, least one test with \trans{ANY, FAILED, PASSED} transition is required.
    \item \textbf{No Regressions:} No test may exhibit a transition that indicates the fix introduced a new bug, so transitions in \trans{PASSED, PASSED, FAILED} and \trans{SKIPPED, SKIPPED, FAILED} are not allowed.
    \item \textbf{Unambiguous Behavior:} Instances with flaky or abnormal transitions are discarded.
\end{enumerate}
Only instances that pass this rigorous validation are included in our final dataset.

\subsection{Framework Analysis}\label{ssec:framework_analysis}
To assess the effectiveness and fidelity of \swemirror{}, we conducted a detailed analysis of our framework. Our goal was to answer three core questions: (1) How effective is our LM-based pre-filter? (2) What is the end-to-end mirror success rate for promising candidates? (3) Are the final mirrored tasks semantically consistent with the original issues and seems realistic?

\paragraph{Effectiveness of LM-based Pre-filter.}
A critical component of our framework's efficiency is the LM-based heuristic, which acts as an intelligent filter to identify \textit{high-quality} and \textit{mirrorable} tasks in \Cref{ssec:task_collection}. To rigorously evaluate its performance, we constructed a balanced evaluation set of 100 issues manually select from issues after the rule-based filtering. This set contains 50 positive instances, which are high-quality and mirrorable, and 50 negative instances, comprising issues that are either low-quality or impossible to mirror. The filter's task is to accept the positive instances while rejecting the negative ones.
As \Cref{tab:filter_confusion} shown, the filter demonstrates a high precision of 84.3\%. This ensures that the vast majority of issues passed to the expensive downstream stages are indeed valuable candidates, thus minimizing wasted computation. Furthermore, with a recall of 86.0\% , the filter successfully captures a large portion of the usable issues. 

\begin{table}[htbp!]
\centering
\begin{tabular}{lcc}
\toprule
& \textbf{Accepted} & \textbf{Rejected} \\
\midrule
Positive & 43  & 7 \\
Negative & 8   & 42 \\
\bottomrule
\end{tabular}
\caption{Confusion matrix for the LM-based filter.}
\label{tab:filter_confusion}
\end{table}

\begin{table}[htbp!]
\centering
\begin{tabular}{lccc}
\toprule
\multirow{2}{*}{\textbf{Language}} & \multirow{2}{*}{\textbf{Yield Rate (\%)}} & \multicolumn{2}{c}{\textbf{Error(\%)}} \\
\cmidrule(lr){3-4}
& & \textbf{Compile/Syntax} & \textbf{Semantic} \\
\midrule
Python & 68.0 & 2.0 & 30.0 \\
Rust & 28.0 & 36.0 & 36.0 \\
Go & 36.0 & 28.0 & 36.0 \\
JavaScript & 52.0 & 6.0 & 42.0 \\
\midrule
\textbf{Overall} & \textbf{46.0} & \textbf{18.0} & \textbf{36.0} \\
\bottomrule
\end{tabular}
\caption{Detailed breakdown of outcomes from the task mirroring phase, with error types categorized.}
\label{tab:mirroring}
\end{table}

\paragraph{Effectiveness of Mirroring.}\label{para:mirror_effectiveness}
We next evaluate the core of our framework: the task mirroring engine. The goal here is to measure the success rate when the pipeline is provided with ideal inputs. For this experiment, we manually select each 100 issues for Python, Rust, Go and Javascript following the same criteria as previous experiment. Result is considered success if it passed the validation in \Cref{ssec:issue_verify}. To gain deeper insight into the failure modes, we further categorized each unsuccessful attempt into one of two types. The first is \textit{Compile/Syntax Error}, which we define as any instance where no tests could be run, typically because the generated patch prevents the project from building or leads to a fatal syntax error. The second is \textit{Semantic Error}, which encompasses all other failures where the code runs, but does not correctly produce the required "fail-to-pass". The results, presented in \Cref{tab:mirroring}, show an overall yield rate of \textbf{46.0\%}. Performance, however, varies significantly by language. Python achieves the highest success rate at 68.0\%, while compiled languages like Rust (28.0\%) and Go (36.0\%) prove more challenging. The error breakdown reveals why: \textit{Compile/Syntax} errors are the dominant failure mode for Rust and Go, accounting for 36.0\% and 28.0\% of their respective totals. In contrast, this error type is rare for the dynamically-typed Python (2.0\%) and JavaScript (6.0\%).

\paragraph{Faithfulness of Mirroring.}
A high yield rate is only meaningful if the generated tasks are faithful representations of the original problems. A task that passes our validation but does not reflect the source issue's core logic is not a useful addition to a dataset. Therefore, our final analysis evaluates the semantic fidelity of the successfully mirrored tasks. To assess this, the 184 tasks successfully generated in \Cref{para:mirror_effectiveness} were independently audited by three human annotators. They compared each generated task instance against the original GitHub issue and PR pair. The results of this audit were highly encouraging. As shown in \Cref{tab:human-agreement}, a consensus was reached on the vast majority of tasks. Out of the 177 tasks with majority results, 156 tasks (88.1\%) were deemed to have either High or Moderate consistency, providing strong evidence that \textsc{\swemirror{}} succeeds in preserving the semantic essence of real-world software engineering challenges.

\begin{table}[htbp]
\centering
\caption{The human classify results on the semantic faithfulness of 184 mirrored tasks.}
\label{tab:human-agreement}
\begin{tabular}{lcccc}

\toprule
& \multicolumn{2}{c}{\textbf{Agreement Pattern}} & \\
\cmidrule(lr){2-3}
\textbf{Final Classification} & \textbf{Unanimous (3-0)} & \textbf{Majority (2-1)} & \textbf{Total} \\
\midrule
High Consistency & 90 & 25 & 115 \\
Moderate Consistency & 30 & 11 & 41 \\
Inconsistent & 15 & 6 & 21 \\
\midrule
Unclassifiable (No Majority) &  &  & 7 \\
\bottomrule
\end{tabular}
\end{table}

\begin{table}[htbp]
\centering
\begin{tabular}{l r r rr rr}
\toprule
& \multicolumn{1}{c}{\textbf{Repos}} & \multicolumn{1}{c}{\textbf{Instances}} & \multicolumn{2}{c}{\textbf{Fix patches}} & \multicolumn{2}{c}{\textbf{Unit tests}} \\
\cmidrule(r){2-2} \cmidrule(lr){3-3} \cmidrule(lr){4-5} \cmidrule(l){6-7}
Language   & \#Num & \#Num     & \#Hunks & \#Lines   & \#P2P    & \#F2P \\
\midrule
Python     & 31  & 46,820  &  3.0  & 38.5   & 1,025.8 &  31.2 \\
Rust       & 6   & 7,183   &  2.4  & 36.8   & 627.3  &  80.2 \\
Go         & 2   & 4,056   &  3.3  & 42.5   & 107.1  &  7.5  \\
JavaScript & 1   & 2,612   &  2.7  & 36.2   & 216.0  &  33.8 \\
\bottomrule
\end{tabular}
\caption{Dataset stastics of \dataset{}}
\label{tab:dataset-stastics}

\end{table}

\subsection{Dataset Statics and Features}
We apply \textsc{\swemirror{}} on 40 repositories accross 4 language. Since we enable sampling in \Cref{ssec:issue_mirroring}, we can sometimes get more than one mirroring results, we perform deduplication to ensure that every instance have different F2P tests and each \textit{\fixpatch{}} modifies different content of the code base. The final dataset comprises 60,671 validated tasks. \Cref{tab:dataset-stastics} presents a detailed statistical overview of the \textsc{SWE-Mirror-60K}.

\section{Experiments}

In this section, we present a comprehensive empirical evaluation of our approach. We first detail the experimental setup, including our agent framework, data collection process, and post-training methodology. We then present the main results on two challenging benchmarks, demonstrating that our datasets boost the performance of base models. Finally, we conduct in-depth ablation studies to analyze the impact of data scale, training strategies, and the role of multi-lingual data.

\subsection{Experimental Setup}

\paragraph{Agent Scaffolding.}
We selected OpenHands~\citep{wang2025openhands}, an open-source, event-driven platform, as the agent framework for all experiments. OpenHands enables LLM agents to iteratively edit files, execute shell commands, and browse the web within sandboxed containers. This framework is known for establishing strong and reproducible baselines on benchmarks like SWE-Bench. For our experiments, we equipped the agent with 3 tools: \textit{str-replace-editor} for file editing and reading, \textit{execute-bash} for command execution and \textit{finish} to stop and submission. we use MOpenHands\footnote{\url{https://github.com/multi-SWE-Bench/MopenHands}} for languages other than Python.

\paragraph{Agent Trajectory Collection.}
To generate training data, we employed high-performing expert LLMs (\textit{Claude-3.7-Sonnet} and \textit{Claude-4-Sonnet}) to produce agent trajectories on a 15k subset of our \textsc{SWE-Mirror-60K} dataset. For each task, we executed 3 trials with a temperature of 1.0 and a maximum of 100 rounds. A trajectory was considered successful only if it ends with a \textit{finish} action and the set of tests passed after applying the submited patch are a superset of the tests fixed by the ground-truth patch. This rigorous process yielded 6,431 successful trajectories. We combined these with 6,025 trajectories from prior experiments on SWE-rebench~\cite{swerebench}, creating a final post-training dataset of 12,456 trajectories.

\paragraph{Agentic Post-training.}
We use \texttt{Qwen2.5-Coder-Instruct-7B~\cite{qwen2025qwen25technicalreport}} and \texttt{32B} models as our base, resulting in our final models, \texttt{SWE-Mirror-LM-7B} and \texttt{SWE-Mirror-LM-32B}. The models were trained for maximum 3 epochs. We utilized AdamW~\citep{loshchilov2019adamw} optimizer with weight decay of 0.01 and cosine learning rate schedule with warmup ratio of 0.1, peaking at learning rate of 5e-5. Specifically, our loss masking technique ensures that the loss is computed only for valid assistant turns that result in well-formed actions, a strategy we analyze in detail in \Cref{ssec:ablation}. For experiments involving trajectories less than 4k, we set maximum learning rate as 1e-4 and trained 5 epochs using trajectories only from \dataset{}.

\paragraph{Evaluation Benchmarks and Metrics}
We evaluate our models on two primary benchmarks. The first, SWE-Bench-Verified \cite{jimenez2024swebench, openai2024swebenchverified}, is a high-quality, human-curated set of 500 real-world software engineering issues in Python. The second, Multi-SWE-Bench-Flash \cite{zan2025msb}, is a benchmark of 300 tasks designed for rapid evaluation of multi-lingual generalization capabilities. Performance is measured by the \textit{Resolved Rate (\%)}, which is the percentage of tasks solved successfully. Key hyperparameters were set as follows: the inference temperature was fixed at 0 for all experiments. The models were trained using a context length of 32,768. For evaluation our model in \Cref{tab:main_results}, we extended the context length to 131,072 with yarn and allowed for a maximum of 100 interaction rounds. For the ablation studies, we used a context length of 32,768 and a maximum of 100 rounds, but keep the model's only the last 5 observations' content from environment in the context.

\subsection{Experiment Results}
Our main experimental results presented in \Cref{tab:main_results} demonstrate the effectiveness of our approach. On the challenging SWE-Bench-Verified benchmark, our \texttt{SWE-Mirror-LM-32B} achieves a resolve rate of \textbf{52.2\%}, matching the performance of much larger models like \texttt{DeepSeek-R1} and \texttt{GPT-4.1} under the same agent framework. Furthermore, on Multi-SWE-Bench-Flash our \texttt{SWE-Mirror-LM-32B} achieves score of \textbf{21.33\%}, outperforming both \texttt{DeepSeek-R1} and \texttt{GPT-4.1}. These results validate that training on a large-scale dataset of mirrored, real-world issues significantly enhances an model's abilities on SWE tasks.
 
\begin{table}[htbp]
\centering
\caption{Performance on SWE-Bench-Verified and Multi-SWE-Bench-Flash (MSB-Flash). The primary metric is \textit{Resolved Rate (\%)}. (V) denotes the Verified subset. MOpenHands\tablefootnote{\url{https://github.com/multi-swe-bench/MopenHands}} refers to the multi language version of OpenHands.}
\label{tab:main_results}
\begin{tabular}{l l c c}
\toprule
\textbf{Model / Method} & \textbf{Scaffold} & {\textbf{SWE-Bench (V)}} & {\textbf{MSB-Flash}} \\
\midrule
\multicolumn{4}{l}{\textit{Proprietary Models}} \\
GPT-4.1-0414~\citep{gpt4.1}                     & (M)OpenHands & 57.6 & 14.33 \\
Claude-4-Sonnet~\citep{anthropicClaudeSonnet}   & SWE-Agent    & 66.6 & {--} \\
                                                & (M)OpenHands & 70.4 & 25.00 \\
\midrule
\multicolumn{4}{l}{\textit{Open-Source Models}} \\
Qwen2.5-Coder-Instruct-7B~\citep{yang2025qwen3technicalreport} & (M)OpenHands &  1.0 & 0.33 \\
SWE-agent-LM-7B~\citep{yang2025swesmith}                       & SWE-Agent    & 15.2 & {--}  \\
\addlinespace
Qwen2.5-Coder-Instruct-32B~\citep{yang2025qwen3technicalreport}& (M)OpenHands &  6.2 & 0.67 \\
SWE-gym-32B~\citep{pan2025swegym}                              & OpenHands    & 20.6 & {--}  \\
SWE-agent-LM-32B~\citep{yang2025swesmith}                      & SWE-Agent    & 40.2 & {--}  \\
DeepSWE-32B-Preview~\citep{deepswe}                            & OpenHands    & 42.2 & {--}  \\
Skywork-SWE-32B~\citep{zeng2025skyworksweunveilingdatascaling}           & OpenHands                          & 47.9 & {--}  \\
\addlinespace
SWE-fixer-72B~\citep{xie2025swefixer}             & SWE-Fixer                          & 32.8 & {--}  \\
Lingma-SWE-GPT-72B~\citep{ma2024lingmaswegpt}        & SWE-Syninfer                       & 32.8 & {--}  \\
\addlinespace

DeepSeek-R1-0528~\citep{deepseekai2025deepseekr1}          & (M)OpenHands & 45.6 & 15.33 \\
Qwen3-Coder-480B-A35B-Instruct & (M)OpenHands & 69.6 & 27.00 \\
\midrule
\multicolumn{4}{l}{\textit{Ours}} \\
\textbf{SWE-Mirror-LM-7B}    & (M)OpenHands & \bfseries 22.8 & \bfseries 6.33  \\
\textbf{SWE-Mirror-LM-32B}   & (M)OpenHands & \bfseries 52.2 & \bfseries 21.33 \\
\bottomrule
\end{tabular}
\end{table}

\vspace{-0.5em}

\begin{figure}[htbp]
    \centering 
    \begin{subfigure}[b]{0.48\textwidth}
        \centering
        \includegraphics[width=\linewidth]{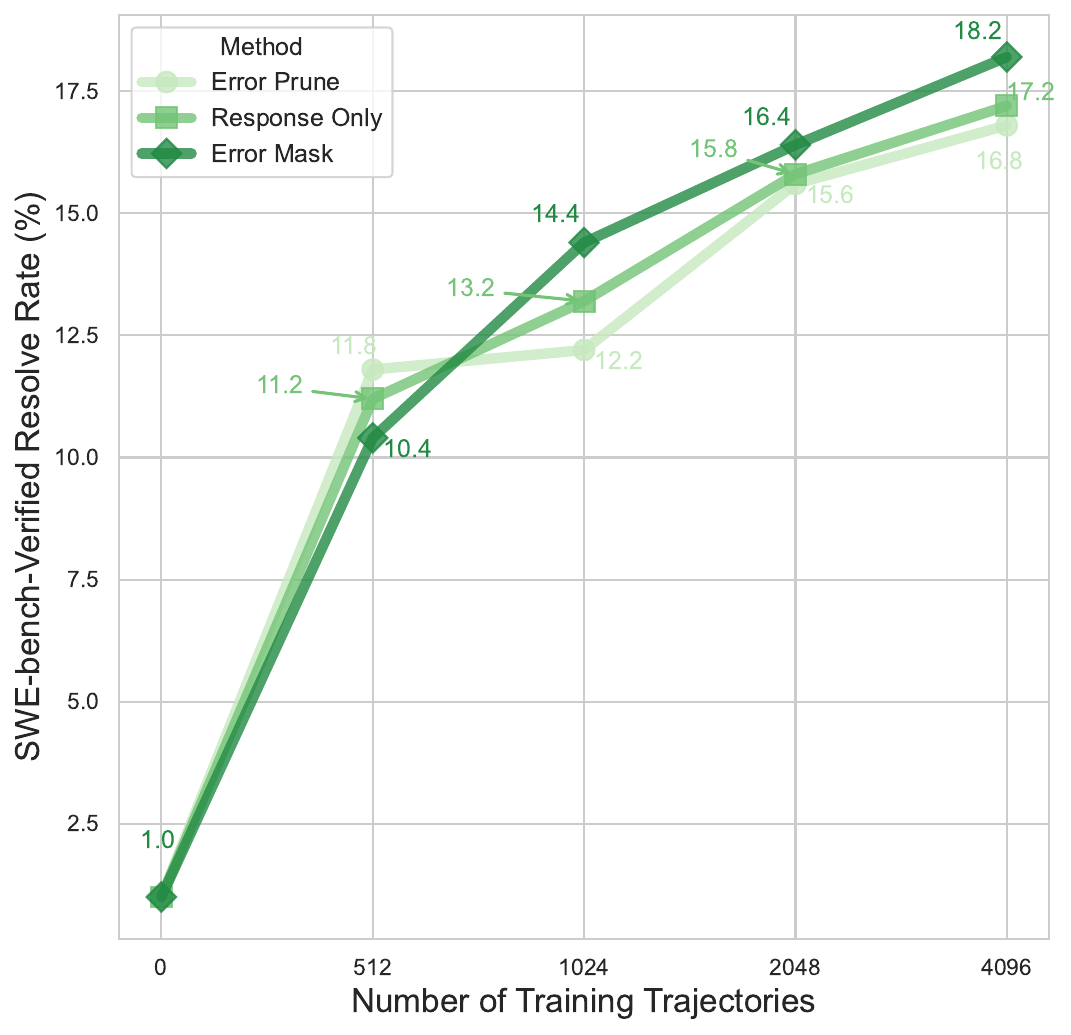}
        \caption{Scaling results for the 7B model.}
        \label{fig:scaling_7B}
    \end{subfigure}
    \hfill 
    \begin{subfigure}[b]{0.48\textwidth}
        \centering
        \includegraphics[width=\linewidth]{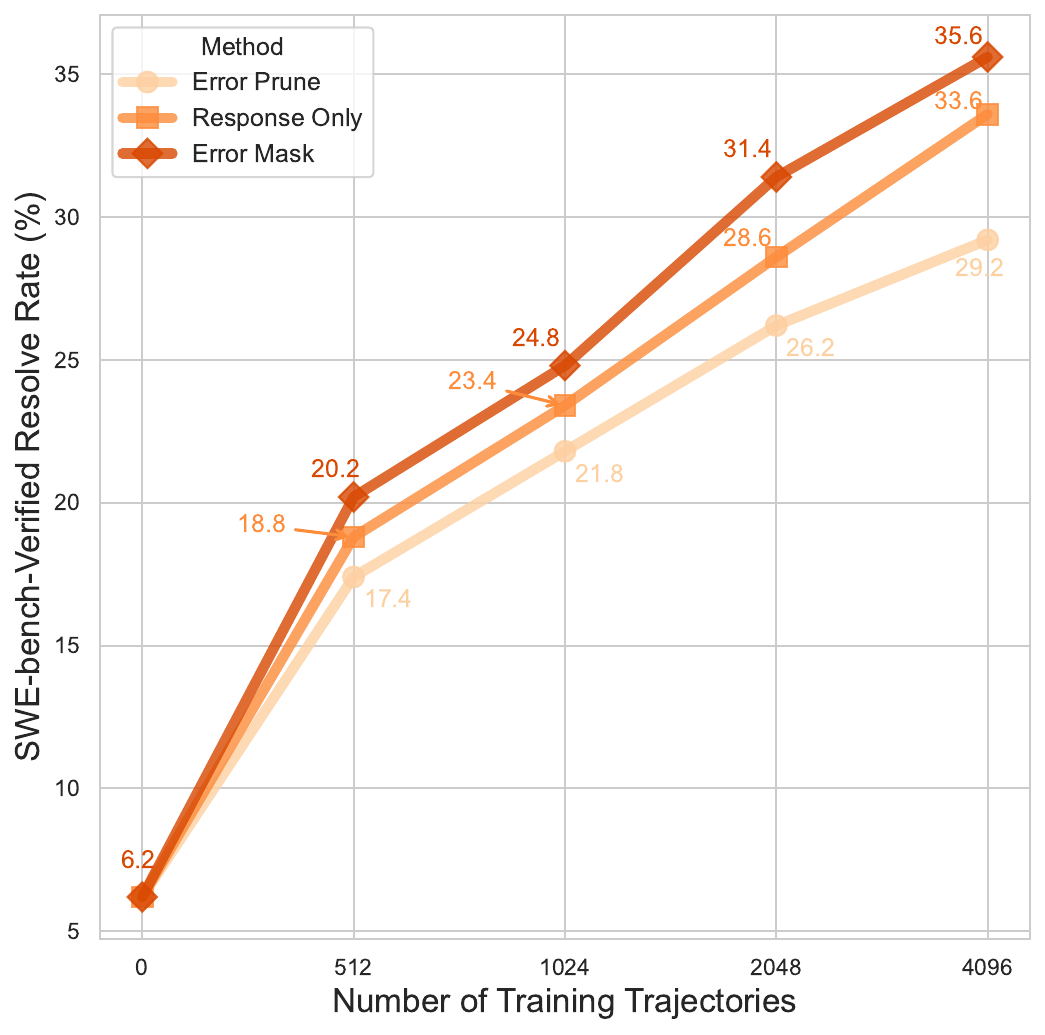}
        \caption{Scaling results for the 32B model.}
        \label{fig:scaling_32B}
    \end{subfigure}
    \caption{Performance on SWE-Bench-Verified as a function of training data scale for our three different training strategies. The \textit{Error Masking} approach consistently outperforms the other methods for both 7B and 32B models.}
    \label{fig:scaling}
\end{figure}

\subsection{Ablation Studies}\label{ssec:ablation}

To dissect the key components contributing to our model's performance, we conduct a series of ablation studies designed to answer two fundamental questions. First, \textit{what are the effects of data scale and the training strategy used to handle errors within demonstration trajectories}? Second, \textit{does training enable the model to generalize across programming languages}? These experiments validate our core design choices regarding the dataset and training methodology and offer valuable insights for future work in agentic post-training.

\subsubsection{Impact of Data Scale and Training Strategy}
A fundamental challenge in training agents from demonstrations is how to handle intermediate error steps within otherwise successful trajectories. Expert-generated trajectories are not always monotonic paths to success; they often contain erroneous actions (\textit{e.g.}, invalid function calls, incorrect arguments) that the expert subsequently self-corrects. Our guiding hypothesis is that training should focus gradient updates on generating valid, productive actions rather than replicating an expert's mistakes. This approach should not only prevent the model from learning to make errors but also improve its ability to recover from them.

To systematically answer this question, we designed and compared three strategies, each embodying a different hypothesis about the role of errors in learning:

\begin{itemize}
    \item \textbf{Response Only:} This standard approach fine-tunes the model on all expert responses, including those that lead to errors. It risks teaching the model to replicate the expert's mistakes.
    \item \textbf{Error Pruning:} This strategy posits that error steps are detrimental and removes any turn containing an error. While this avoids reinforcing mistakes, it comes at the high cost of discarding the context of how an agent recovers from an error, thereby losing learning opportunity for self-correction.
    \item \textbf{Error Masking:} This strategy, which embodies our central hypothesis, preserves the full trajectory context but surgically masks the loss on erroneous agent responses. This allows the model to learn from the context of a mistake without learning to make the mistake. By applying all gradient updates to valid actions, this method provides a rich learning signal for both action generation and error recovery.
\end{itemize}

\Cref{fig:scaling} plots the resolve rate on SWE-Bench-Verified as a function of the number of training trajectories from \dataset{}. The results validate the quality of our dataset and reveal two clear trends:

\begin{enumerate}
    \item \textbf{A Strong Scaling Law with Data:} For both model sizes and across all strategies, performance consistently improves as the number of training data increases~\citep{kaplan2020scalinglaw}. The 32B model trained with our \textit{Error Masking} strategy improves its resolve rate from a baseline of 6.2\% to 35.6\% when trained on 4096 trajectories. This demonstrates a direct and powerful correlation between data volume and issue-resolving capability.

    \item \textbf{The Superiority of Error Masking:} The \textit{Error Masking} strategy consistently outperforms both \textit{Error Pruning} and the baseline across all data scales and model sizes. The performance gap between \textit{Error Masking} and the other methods widens as the dataset grows, suggesting that the benefits of its richer learning signal compound with more data. By observing the entire sequence, the model learns how to recover from error states—a crucial skill that is lost when imperfect data is pruned. This makes \textit{Error Masking} a more data-efficient and effective approach for training robust models.
\end{enumerate}

\subsubsection{Cross-Lingual Generalization}
To quantify the benefit of our multi-lingual dataset, we evaluated whether non-Python data could improve performance on the Python-only SWE-Bench Verified benchmark. We trained the 7B model on several monolingual subsets of our data (512 trajectories each) using the \textit{Error Masking} recipe.

The results, presented in \Cref{tab:non_python}, are compelling. The model trained exclusively on non-Python data still achieves a notable resolve rate on Python tasks. This provides strong evidence of \textbf{cross-lingual generalization}, wherein the model learns abstract problem-solving patterns and code semantics that transfer across languages. Notably, the model trained on Rust data yielded the most significant performance gain, which we attribute to the language's complexity and rich type system fostering more robust reasoning capabilities.

\begin{table}[htbp]
\centering
\begin{tabular}{lcl}
\toprule
\textbf{Training Language} & \textbf{Resolve Rate (\%)} & \textbf{Improvement (\%)} \\
\midrule
Base model             & 1.0                  & --                  \\
+ Go                   & 10.2                 & $\uparrow 9.2$                \\
+ Rust                 & 11.3                 & $\uparrow 10.3$               \\
+ JavaScript           & 9.4                  & $\uparrow 8.4$                \\
\bottomrule
\end{tabular}
\caption{Cross-lingual generalization performance on the Python SWE-Bench Verified benchmark.}
\label{tab:non_python}

\end{table}
\section{Related Work}
\paragraph{Coding Agents.}
Recent advancements in Software Engineering have spurred the development of agents capable of resolving real-world issues in repositories. These agents are evaluated on benchmarks like SWE-bench~\citep{jimenez2024swebench} and Multi-SWE-bench~\citep{zan2025msb}. A significant body of work focuses on agent design. For instance, OpenHands~\citep{wang2025openhands} introduces an event-driven platform that empowers LLM agents to iteratively edit files and execute commands. SWE-Agent~\cite{yang2024sweagent} introduces Agent-Computer Interface (ACI) to provide LLM agents with actions for operating computer like editors and shells. In contrast to relying on an LLM's autonomous decision-making, another line of research argues for utilizing structured workflow architectures. Agentless~\citep{xia2024agentless}, Agentless-Mini~\citep{wei2025swerl} and Moatless~\citep{moatless} demonstrate that combining workflow with test-time scaling can outperform many sophisticated SWE agents on SWE-bench while reducing computational costs. Some research works also explored the self-evolution of coding agents, exemplified by GDM~\citep{zhang2025dgm}, SE-Agent~\citep{lin2025se} and SWE-Exp~\citep{chen2025sweexp}, showing impressive improvement. Another research area has focused on enhancing the models themselves. SWE-Fixer~\cite{xie2025swefixer} represents a learning-based approach, improving file retrieval and patch generation capabilities using supervised fine-tuning to effectively train open-source LLMs for specialized SWE tasks. SWE-Gym~\citep{pan2025swegym} and SWE-Smith~\citep{yang2025swesmith} have explored rejection sampling fine-tuning, an approach that we also adopt in our work. Furthermore, reinforcement learning (RL) has been utilized to refine model capabilities, with SWE-RL~\citep{wei2025swerl} using patch similarity as a reward signal and SWE-Swiss~\citep{sweswiss}, DeepSWE~\citep{deepswe} and SkyRL~\citep{cao2025skyrl} exploring execution-based rewards as a promising future direction.

\paragraph{Issue-Resolving Datasets.}
The development of datasets for training and evaluating issue-resolving agents has rapidly progressed from static code collections to dynamic, interactive environments. A foundational contribution is SWE-Gym~\citep{pan2025swegym}, which established the paradigm of using real-world Python issues paired with executable environments and unit tests, enabling interactive agent training and verification. To combat the growing problem of data contamination in static benchmarks, SWE-rebench~\citep{swerebench} and  SWE-Factory~\citep{guo2025swefactory} introduced a dynamic pipeline that continuously sources fresh, decontaminated tasks from active GitHub repositories, ensuring a more robust and reliable evaluation of an agent's true generalization capabilities. Recognizing that manual curation remains a significant bottleneck, subsequent efforts have focused on scalable, automated data generation. SWE-Smith~\cite{yang2025swesmith} pioneered a synthetic approach by inverting the typical workflow, starting with working code and automatically injecting bugs to create thousands of new tasks. Similarly, SWE-Synth~\cite{pham2025swesynth} uses LLMs to simulate the entire debugging process, generating not just code fixes but also test cases and structured repair trajectories. Complementing these, R2E-Gym~\cite{jain2025r2e} leverages a procedural generation pipeline to curate large-scale training environments directly from code commits, reducing the reliance on human-written issues. Together, these works highlight a critical trend towards creating more scalable, realistic, and verifiable data sources to advance agentic software engineering.

\section{Conlusion}
This paper introduces \swemirror{}, a novel pipeline which multiplies the utility of each Gym and unlocks the vast history of software evolution on platforms like GitHub as a source of training data. Our primary contribution is the release of SWE-Mirror-60K, a large-scale dataset of 60,000 verifiable tasks built using this methodology. Our empirical evaluations demonstrated that models finetuned on \dataset{} exhibit significant improvements in their issue-resolving capabilities, validating the quality and effectiveness of our approach. Furthermore, our in-depth ablation studies provide critical insights for the field. We have also confirmed a strong scaling law where performance consistently improves with data volume, demonstrated the efficiency of \textit{Error Masking} training strategy and revealed the evidence of cross-lingual generalizability, where models trained exclusively on non-Python data still exhibit notable proficiency on Python tasks, highlighting the value of multi-lingual data in learning generalized, abstract problem-solving patterns.


\clearpage

\bibliographystyle{plainnat}
\bibliography{main}

\clearpage

\beginappendix

\section{Appendix}

\subsection{Pull-Request Collection and Filter}\label{appdix:pr_collection}
We use following rules to collect high-quality pull-requests:
\begin{itemize}
    \item It must have linked issues;
    \item It must have been merged and closed;
    \item It must edit code files.
\end{itemize}
Unlike SWE-bench~\citep{jimenez2024swebench}, our filtering criteria do not require pull requests to modify test files. This is for two reasons: first, it is difficult to isolate test modifications in languages like Rust where tests are co-located with source code; second, we generate tests separately using a \textit{Test Agent}. To finalize our dataset, we use an LLM for quality control and to predict mirrorability, as guided by the following prompt.

\begin{tcolorbox}[notitle, sharp corners, breakable, colframe=Periwinkle, colback=white, 
       boxrule=3pt, boxsep=0.5pt, enhanced, 
       title={Prompt for LLM Filter},]\label{box:gradient}
       {\scriptsize
\begin{lstlisting}
prompt = """You are a senior software engineer. 

You are given a pull request from another repository.

You are going to check, response True in final answer if the pull request is a bug fix or a feature addition, and response False if the pull request is just fixing some error messages or documentations.
1. is the pull request a bug fix or a feature addition
2. is the pull request non-trivial, just fixing error messages, docs, also, this not-related to external dependencies.
3. if some functionality related to the bug or feature exists in the current repository.

Belowing is the description of the pull request:
<pull_request>
   <body>
   {body}
   </body>
   <diff>
   {diff}
   </diff>
</pull_request>

Belowing is the readme and the test suite of the current repository:
<current_repo>
   <readme>
   {readme}
   </readme>
   <test_suite>
   {test_suite}
   </test_suite>
</current_repo>


Think Step by Step with following questions
1. What is the bug fixed or the feature added in the pull request?
2. What is the related functionality of the bug?
3. Does the current repository have the related functionality:
    1. If yes, what is the related functionality?
4. Is it possible to introduce the bug/feature in the current repository?

Note:
- The language of repos does not matter, you should focus on the functionality of the bug.

Respond with python list with two elements, "exists", "reason", in the following format:
```python
[True/False, "The pull request is a bug fix or a feature addition, related to ...., the current repository has the related functionality."] 
```
"""
\end{lstlisting}
}
\end{tcolorbox}

\subsection{Task Mirroring Workflow}\label{appdix:workflow}
The first step is to distill the core symptoms and logic from pull requests in similar repositories. We introduced this step for a critical reason: raw pull request and issue descriptions often contain repository-specific information (\textit{e.g.}, variable names, file paths, and stack traces). This context-specific data can mislead the model into localizing non-existent files or generating patches that result in compilation or syntax errors. The distillation process, therefore, focuses on extracting the underlying functionality, core logic, current and expected behavior, and observable symptoms. The prompts used for this task are provided below.
\begin{tcolorbox}[notitle, sharp corners, breakable, colframe=Periwinkle, colback=white, 
       boxrule=3pt, boxsep=0.5pt, enhanced, 
       title={Prompt for Problem Abstraction},]\label{box:gradient}
       {\scriptsize
\begin{lstlisting}
Consider the following pull request that fixes a bug:
<pull_request>
   <body>
   {body}
   </body>
   <diff>
   {diff}
   </diff>
</pull_request>

Your task is to abstract the bug pattern from the pull request, focusing exclusively on systemic issues that require changes in multiple locations across the codebase.

Here is an example of a complex bug pattern that requires multiple edits:
<pull_request>
   <body>
   Fix inconsistent error handling across API endpoints

   Multiple API endpoints were handling validation errors differently, leading to inconsistent error responses and poor user experience. Some endpoints returned 400 status codes while others returned 500, and error message formats varied. This PR standardizes error handling across all user-facing endpoints to provide consistent behavior.

   The fix involves:
   - Updating user registration endpoint error handling
   - Fixing profile update validation responses  
   - Standardizing login error messages
   - Adding consistent error formatting in shared utilities

   Fixes #456
   </body>
   <diff>
   @@ -8,7 +8,8 @@ class UserController:
        def register(self, user_data):
            if not self.validate_user_data(user_data):
   -            return {{"error": "Bad input"}}, 500
   +            return {{"error": "Invalid user data", "details": self.get_validation_errors(user_data)}}, 400
   
   @@ -22,7 +23,8 @@ class UserController:
        def update_profile(self, user_id, profile_data):
            if not self.validate_profile_data(profile_data):
   -            raise Exception("Validation failed")
   +            return {{"error": "Invalid profile data", "details": self.get_validation_errors(profile_data)}}, 400
   
   @@ -35,6 +37,7 @@ class AuthController:
        def login(self, credentials):
            if not self.validate_credentials(credentials):
   -            return {{"message": "Login failed"}}, 500
   +            return {{"error": "Invalid credentials", "details": "Username or password incorrect"}}, 401
   
   @@ -5,6 +5,10 @@ class ValidationUtils:
   +    def get_validation_errors(self, data):
   +        # Standardized error formatting
   +        return [str(error) for error in self.validator.errors(data)]
   +
        def validate_user_data(self, data):
            return self.validator.is_valid(data)
   </diff>
</pull_request>

Follow this pattern when abstracting the bug - identify systemic issues that manifest across multiple files and functions:

```md
### Bug Pattern

**Issue Type**: Inconsistent Error Handling / API Response Standardization

**Core Problem**: 
The application lacks consistent error handling patterns across similar functions or modules, leading to unpredictable behavior and poor user experience. Different parts of the codebase handle similar error conditions in incompatible ways.

**Technical Context**:
- API endpoints or service methods that perform similar validation or processing
- Error handling logic scattered across multiple controllers, services, or utility functions
- Inconsistent status codes, error message formats, or exception handling approaches
- Missing standardized error response structures

**Symptom**:
- Different error responses for similar failure conditions
- Inconsistent HTTP status codes across related endpoints
- Varying error message formats that confuse API consumers
- Some functions throw exceptions while others return error objects

**Root Cause Pattern**:
- Lack of centralized error handling utilities or standards
- Copy-paste development without following established patterns
- Missing shared validation or error formatting functions
- Inconsistent exception handling strategies

**Impact Scope**:
Multiple locations typically affected:
- All API endpoints that perform user input validation
- Service layer methods that process similar data types
- Controller functions handling authentication or authorization
- Utility functions used for data processing or validation
- Error response formatting across different modules
```

Please wrap the bug pattern in the following format:
```md
.. the bug pattern ..
```
"""

\end{lstlisting}
}
\end{tcolorbox}
For \textit{Test Agent} and \textit{Mirror Agent}, we implement them in Agentless style, each go through: (1) localize related file and (2) genearte patch in \textit{Search/Replace} format.

\begin{tcolorbox}[notitle, sharp corners, breakable, colframe=Periwinkle, colback=white, 
       boxrule=3pt, boxsep=0.5pt, enhanced, 
       title={Test Agent: Prompt for Localization},]\label{box:gradient}
       {\scriptsize
\begin{lstlisting}
TEST_LOCALIZE = """\
Please look through a given issue description and repository structure and provide two list of files related to the issue:
- `source_files`: the files may contains code related to the functionality described in the issue
- `test_files`: the files which should contain the test cases for the functionality described in the issue

--- BEGIN ISSUE ---
{issue}
--- END ISSUE ---

--- BEGIN REPOSITORY STRUCTURE ---
{structure}
--- END REPOSITORY STRUCTURE ---

Only provide the full path and return at most {n} files for each list. 

Respond in the following format, wrapped your results in a markdown python code block with a dictionary with two keys `source_files` and `test_files`.
```python
{{
    "source_files": [
        "most/important/file1.xx",
        "less/important/file2.yy",
        ...
    ],
    "test_files": [
        "most/important/file1.xx",
        "less/important/file2.yy",
        ...
    ]
}}
```
\end{lstlisting}
}
\end{tcolorbox}
\begin{tcolorbox}[notitle, sharp corners, breakable, colframe=Periwinkle, colback=white, 
       boxrule=3pt, boxsep=0.5pt, enhanced, 
       title={Test Agent: Prompt for Patch Generalization},]\label{box:gradient}
       {\scriptsize
\begin{lstlisting}
TEST_PATCHGEN = """We are currently adding unit tests to the avoid the future regression for functionality described in the issue.

--- BEGIN ISSUE ---
{issue}
--- END ISSUE ---

Below are some source code segments related to the functionality described in the issue.

--- BEGIN SOURCE FILES ---
{source_files}
--- END SOURCE FILES ---

Below are some files you can edit to add unit tests.
--- BEGIN TEST FILES ---
{test_files}
--- END TEST FILES ---

Please first localize the code in SOURCE FILES to the functionality described in the issue and \
then generate *SEARCH/REPLACE* edits to test to some of TEST FILES to test the issue.

Every *SEARCH/REPLACE* edit must use this format:
1. The file path
2. The start of search block: <<<<<<< SEARCH
3. A contiguous chunk of lines to search for in the existing source code
4. The dividing line: =======
5. The lines to replace into the source code
6. The end of the replace block: >>>>>>> REPLACE

Here is an example:

```
{diff_example}
```

Please note that the *SEARCH/REPLACE* edit REQUIRES PROPER INDENTATION. If you would like to add the line '        print(x)', you must fully write that out, with all those spaces before the code!
Wrap each *SEARCH/REPLACE* edit in a code block as shown in the example above. If you have multiple *SEARCH/REPLACE* edits, use a separate code block for each one.

Please make sure the tests you add are not too simple and can be passed by the existing code.
"""

\end{lstlisting}
}
\end{tcolorbox}
\begin{tcolorbox}[notitle, sharp corners, breakable, colframe=Periwinkle, colback=white, 
       boxrule=3pt, boxsep=0.5pt, enhanced, 
       title={ Mirror Agent: Prompt for Localization},]\label{box:gradient}
       {\scriptsize
\begin{lstlisting}
MIRROR_LOCALIZE = """\
Please look through a given issue description, repository structure, a patch related to test the issue and provide a list of files related to the issue

Below is the issue description and repository structure.
--- BEGIN ISSUE ---
{issue}
--- END ISSUE ---

Below is the repository structure.
--- BEGIN REPOSITORY STRUCTURE ---
{structure}
--- END REPOSITORY STRUCTURE ---

Below is the patch applied to the repository to test the issue.
--- BEGIN TEST PATCH ---
{testgen_patch}
--- END TEST PATCH ---

Only provide the full path and return at most {n} files. 

Respond in the following format, wrapped your results in a markdown python code block with a list of files.
```python
[
    "most/important/file1.xx",
    "less/important/file2.yy",
    ...
]
```

""".strip()
\end{lstlisting}
}
\end{tcolorbox}
\begin{tcolorbox}[notitle, sharp corners, breakable, colframe=Periwinkle, colback=white, 
       boxrule=3pt, boxsep=0.5pt, enhanced, 
       title={Mirror Agent: Prompt for Patch Generalization},]\label{box:gradient}
       {\scriptsize
\begin{lstlisting}
MIRROR_PATCHGEN = """We are currently implementing the issue described in the following issue description.

--- BEGIN ISSUE ---
{issue}
--- END ISSUE ---

Below are some code segments related to the issue.

--- BEGIN FILES---
{files}
--- END FILES---

Below is the patch applied to the repository to test the issue, please DO NOT modify any test code or test files.
--- BEGIN TEST PATCH ---
{testgen_patch}
--- END TEST PATCH ---


Here is the list of testcases related to the issue.
--- BEGIN TESTS ---
{tests}
--- END TESTS ---


Please first localize the related source code based on the issue description, and then generate *SEARCH/REPLACE* edits to re-implement the issue via breaking the tests in the TESTS section.
DO NOT modify any test code or test files, you should only modify the non-test files and code related to the issue.

Every *SEARCH/REPLACE* edit must use this format:
1. The file path
2. The start of search block: <<<<<<< SEARCH
3. A contiguous chunk of lines to search for in the existing source code
4. The dividing line: =======
5. The lines to replace into the source code
6. The end of the replace block: >>>>>>> REPLACE



Here is an example:

```
{diff_example}
```

Please note that the *SEARCH/REPLACE* edit REQUIRES PROPER INDENTATION. If you would like to add the line '        print(x)', you must fully write that out, with all those spaces before the code!
Wrap each *SEARCH/REPLACE* edit in a code block as shown in the example above. If you have multiple *SEARCH/REPLACE* edits, use a separate code block for each one.
"""
\end{lstlisting}
}
\end{tcolorbox}

\section{Contributors}
\begin{itemize}
    \item \textbf{Junhao Wang:} Designed and implemented the core framework, including the task mirroring engine and the data processing pipeline, wrote the the manuscript.
    \item \textbf{Daoguang Zan:} Served as Project Leader. He conceived the core idea, directed the overall research, coordinated the team, and contributed to the revision of the manuscript.
    \item \textbf{Shulin Xin:} Conducted the experimental evaluations by executing the models on all benchmarks and was responsible for compiling the results.
    \item \textbf{Siyao Liu:} Managed the project's infrastructure and integrated the data synthesis pipeline with the remote container environment.
    \item \textbf{Yurong Wu:} Contributed to the implementation of the various model training strategies.
    \item \textbf{Kai Shen:} Provided supervision for the project.
\end{itemize}

\end{document}